# Energy-Efficient Distributed Processing in Vehicular Cloud Architecture

Fatemah S. Behbehani, Mohamed Musa, Taisir Elgorashi, and J. M. H. Elmirghani
*School of Electronic and Electrical Engineering, University of Leeds, Leeds, LS2 9JT, United kingdom*
*Email:{elfsma@leeds.ac.uk,M.Musa@leeds.ac.uk,T.E.H.Elgorashi@leeds.ac.uk,J.M.H.Elmirghani@leeds.ac.uk }*

**ABSTRACT**

Facilitating the revolution for smarter cities, vehicles are getting smarter and equipped with more resources to go beyond transportation functionality. On-Board Units (OBU) are efficient computers inside vehicles that serve safety and non-safety based applications. However, much of these resources are underutilised. On the other hand, more users are relying now on cloud computing which is becoming costly and energy consuming. In this paper, we develop a Mixed Integer linear Programming (MILP) model that optimizes the allocation of processing demands in an architecture that encompasses the vehicles, edge and cloud computing with the objective of minimizing power consumption. The results show power savings of 70%-90% compared to conventional clouds for small demands. For medium and large demand sizes, the results show 20%-30% power saving as the cloud was used partially due to capacity limitations on the vehicular and edge nodes.

**Keywords**: Distributed Processing, Wireless, Edge Nodes, Vehicular Networks,

## 1. INTROUCTION

End users are growing more dependent on cloud services and data centers [1]. As the demand on cloud services grows higher, the data centers, as expected, tend to grow even bigger and more expensive in term of both monetary cost and energy consumptions. The energy consumption of clouds and data centers is contributing much to the total cost and power consumption in the Information and Communication Technology (ICT) field. That is why a lot of effort is being put forward now to explore alternatives that are more energy efficient and still as powerful [2-11]. One approach that is being actively evaluated is distributed service providers or the installation of mini data centers close to end users' level. In [12] data processing is done at different layers of the network and not only in the core cloud through optimized placement of Virtual Machines (VM) in IoT devices. Comparison between centralized data centers and nano data centers, to show the validity of the small data centers and its impacting factors, was carried out in [13]. The work in [14] analysed the energy consumption and latency of computation offloading in mobile clouds.

Modern vehicles are increasingly being viewed as smart machines with plenty of computing resources. Research in the area of vehicular networks is very promising and it varies from Internet of Vehicles (IoV) [15] to Vehicular Clouds to VaaR (vehicle as a Resource) [16]. Our work presents an end-to-end architecture that uses vehicular and edge computing as the first level of processing resources. It compares this architecture with conventional clouds from an energy consumption point of view. For the remainder or the paper, Section 2 presents the proposed architecture. Section 3 discusses the optimization model and its results, and in Section 4 the paper is concluded.

## 2. VEHICULAR DISTRIBUTED COMPUTING ARCHITECTURE

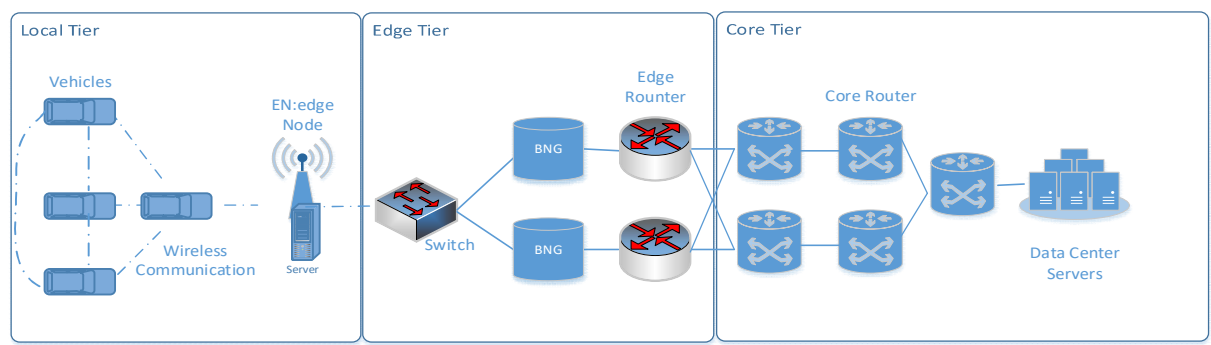

Figure 1: Architecture of distributed processing service using vehicles, edge nodes, and the cloud

The architecture is divided into three layers, as can be seen in Figure 1. The first layer is made of the vehicles. Modern vehicles are now equipped with On Board Units (OBU) of very high computational performance and communication capabilities. They form the first choice for processing destination. The second layer is formed of edge nodes. It provides second level for processing destination close to end user through dedicated servers. Both of these layers are in the local tier of the architecture. The last layer is the conventional cloud servers, which are geographically distant but of powerful computing capabilities. The edge and core switches and routers architecture is based on [17]. The model assumes the vehicles and the edge nodes communicate using one or more wireless interfaces such as Dedicated Short Range Communication (DSRC) and WiFi. It is assumed that the source node (the one that has the request) has knowledge of the available resources of the other nodes, and it chooses where to send its request to be served. This forms distributed type of dynamic control, which can be implemented using Software-Defined Network (SDN) concepts [18, 19].

## 3. OPTIMISATION MODEL AND RESULTS

We developed a Mixed Integer linear Programming (MILP) model to optimise the location in which a demand is served with minimum power consumption. The power consumed in transmission, reception, and processing, increasing linearly with the demand, are calculated for each of the nodes used, in addition to the idle power for each one. The model ensures the conservation of processing capacity of each node, as well as the communication interface bandwidth capacity. Also, the full traffic is sent to every processing node regardless of the assigned processing in it. The model is evaluated in a parking lot with 20 cars parked within 150 meters from each other, communicating using the DSRC interface. The parking lot is surrounded by 4 edge nodes, and each node is connected to 5 vehicles through the WiFi. Each edge node is composed of a server and an access point and they can also communicate with each other using WiFi. Demands are generated by the vehicles and are composed of two parts, the data to be sent (traffic in kbps), and the processing it requires (MIPS). We use the estimations made in [20] for the traffic and the associated processing, where it takes 2000 instructions per bit for the video generation application that is selected as an example for the evaluation. Table 1 shows the parameters values for the three types of processing nodes. Cloud servers are highly efficient with processing efficiency of 4 instructions per cycle [27]. For edge nodes we considered processing efficiency of 3 instructions per cycle. The OBUs in vehicles are efficient enough to accommodate real-time and safety applications with 2 instructions per cycle for the vehicles. We assumed the power consumption in the cloud server and in the edge nodes server (the Raspberry Pi in Table 1) to be consumed fully in processing. The communication power in the edge, on the other hand, is solely consumed by the access point, while for cloud the communication power is the power consumed in the routers and switches leading to it. For the vehicles the same approach in [21] is followed where a typical type of computer is said to dedicate 58% of its power consumption to processing and 21% to communication, and the rest to storage.

*Table 1: performance evaluation parameters*

| Parameter | Values for vehicles | Values for edge nodes | Values for the cloud servers |
|---|---|---|---|
| Max power | 10 W [22] | AP = 25 W, Raspberry Pi = 12.5 W [23, 24] | 301 W [25] |
| Idle power | 5 W [22] | AP = 5.5 W Raspberry Pi = 2 W [23, 24] | 201 W [25] |
| Processor | 800 MHz [22] | 1.2 GHz [24] | 2.5 GHz [25] |
| Processing capacity | 2 * 800 = 1600 MIPS | 3 * 1.2 = 3600 MIPS | Unlimited (10000 per server and assuming multiple servers) |
| DSRC communication capacity | 27 Mb/s | N/A | N/A |
| WiFi communication Capacity | 150 Mb/s [23] | 150 Mb/s [23] | N/A |
| Processing efficiency | 1600/(5*0.58)=550 MIPS/W | 3600/(10.5)=340 MIPS/W | 10000/(100)=100 MIPS/W |

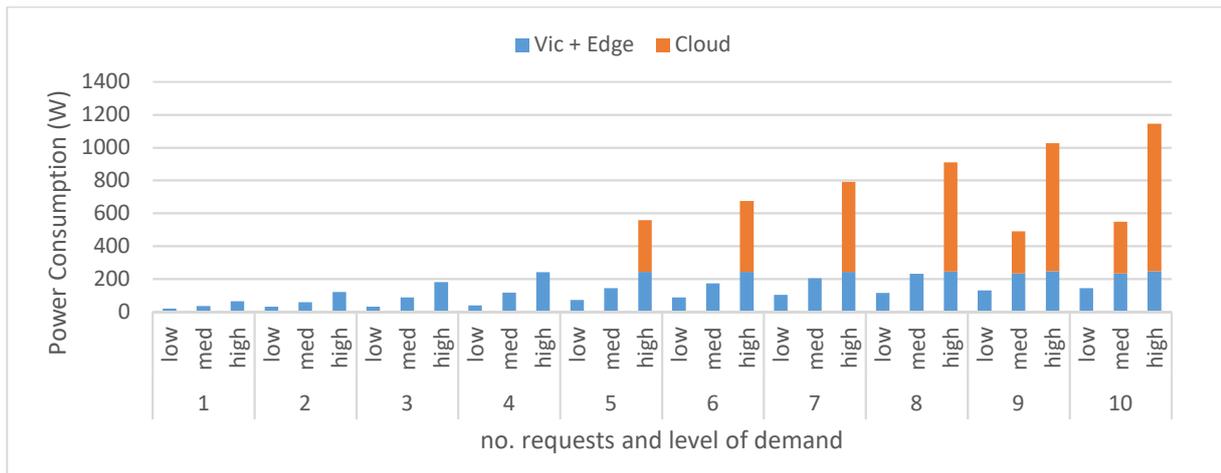

*Figure 2 : Power Consumption for the three demand levels, showing the distribution of the consumption*

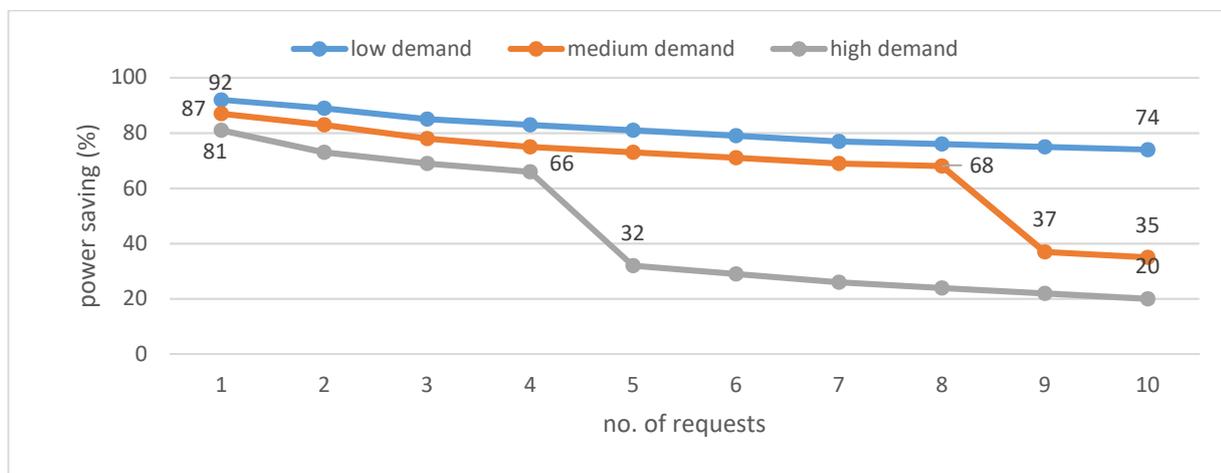

*Figure 3: Power saving of distributed processing in the three types of resources in comparison with conventional Cloud*

The model is evaluated for a varied number of requests ranging from 1 to 10 requests of equal demands initiated by vehicles. The demands are classified into low, medium and high based on the size of the single demand and consequently the total demand. Three demand values representing small video (2880 MIPS), medium video (2 times bigger) and large video (4 times bigger), were used and the traffic was generated according to [20]. The model decides where to serve each and how many nodes to use. We use the term "serve locally" to indicate the use of vehicles and edge node.

**Error! Reference source not found.** shows the total power consumption at the three levels of processing resources available (vehicles, edge nodes, cloud) under the three demand levels as the number of requests increases. For small demands, it is most power efficient to serve them locally at all times. As for medium demand sizes, the demands are served locally if the requester processing demand is within local processing capacity. For 9 and 10 requests, the vehicles and edge nodes were no longer enough to serve the full processing demand, therefore, the part exceeding the vehicles and edge nodes capacities was sent to the cloud, which explains the large increase in the power consumption when compared with 8 requests. The same trend is noticed in the high demands' values but at lower number of requests. The processing capacities of the vehicles and edge nodes were exceeded after 4 requests due the large demand size which creates competition on the use of local resources.

**Error! Reference source not found.** shows the power saving percentage comparing local processing with the conventional cloud. The small demand values experience power saving between 90% for a single request and 74% for 10 requests. For the medium demand values, the power saving is slightly less than the small demands with power saving of 87% for a single request and decreasing to 68% for 8 requests. For the high demands, the power saving starts at 81 % for single requests and drops to 66% for 4 requests. This reduction in power saving is due to the need for more service nodes as the demand increases. The increase in serving nodes means more idle power is used as well as more redundancy in the traffic since each serving node needs to receive the full amount of traffic. For the medium and high demand values, the substantial drop in power saving happens at 9 requests and 5 requests respectively. These are the instances when the model was forced, by limitation of processing capacity of vehicles and edge nodes, to the cloud to be served, causing the power saving to drop to levels of 20%-30%.

## 4. CONCLUSIONS

In this paper, we presented a distributed service architecture based on vehicular network and edge computing for processing demands originating from stationary vehicles. A MILP model was developed to minimise the power consumption by optimising the use of vehicles and edge nodes to perform processing. The performance of the model was evaluated for three processing demand volumes and the results were compared with the use of conventional cloud servers to serve all the demands. The model shows power savings up to 90% for smaller demand values, and a gradual decrease in savings for medium and high demands due to partial use of cloud servers. This saving can be improved by reducing the amount of demand forced into the cloud served, and this in turn can be achieved by increasing processing and communication capacities of vehicles and edge nodes.

## ACKNOWLEDGEMENTS

The authors would like to acknowledge funding from the Engineering and Physical Sciences Research Council (EPSRC), through INTERNET (EP/H040536/1) and STAR (EP/K016873/1) projects. All data are provided in full in the results section of this paper.